\pdfoutput=1
\documentclass[referee]{aa} 

\usepackage{graphicx}
\usepackage{tabularx}
\usepackage{amsmath}
\usepackage[colorlinks=true,linkcolor=magenta,citecolor=blue,filecolor=cyan,urlcolor=purple,final=true]{hyperref}
\usepackage{txfonts}
\begin{document}

   \title{The Automatic Identification and Tracking of Coronal Flux Ropes - Part I: Footpoints and Fluxes}

   \author{A. Wagner
          \inst{1,2},
          E. K. J. Kilpua
          \inst{1}
          \and
          R. Sarkar
          \inst{1}
          \and
          D. J. Price
          \inst{1}
          \and
          A. Kumari
          \inst{1, 4}
          \and
          F. Daei
          \inst{1} 
          \and
          J. Pomoell
          \inst{1}
           \and
          S. Poedts
          \inst{2,3}}

    \institute{Department of Physics, University of Helsinki, P.O. Box 64, FI-00014, Helsinki, Finland\\
    \email{andreas.wagner@helsinki.fi}
    \and
    CmPA/Department of Mathematics, KU Leuven, Celestijnenlaan 200B, 3001 Leuven, Belgium
    \and
    Institute of Physics, University of Maria Curie-Sk{\l}odowska, ul.\ Radziszewskiego 10, 20-031 Lublin, Poland \and
    NASA Goddard Space Flight Center, Greenbelt, MD 20771, USA \\
    }
   \date{Received XXX; accepted XXX}

\abstract
{Investigating the early-stage evolution of an erupting flux rope from the Sun is important to understand the mechanisms of how it looses its stability and its space weather impacts.}
{Our aim is to develop an efficient scheme for tracking the early dynamics of erupting solar flux ropes and use the algorithm to analyse its early-stage properties. The algorithm is tested on a data-driven simulation of an eruption that took place in active region AR12473. We investigate the modelled flux rope's footpoint movement and magnetic flux evolution and compare with observational data from the Solar Dynamics Observatory's (SDO) Atmospheric Imaging Assembly (AIA) in the 211~\AA{} and 1600~\AA{} channels.}
{To carry out our analysis, we use the time-dependent data-driven magnetofrictional model (TMFM). We also perform another modelling run, where we stop the driving of the TMFM midway through the flux rope's rise through the simulation domain and evolve it instead with a zero-beta magnetohydrodynamic (MHD) approach.}
{The developed algorithm successfully extracts a flux rope and its ascend through the simulation domain. We find that the movement of the modelled flux rope footpoints showcases similar trends in both TMFM and relaxation MHD run: they recede from their respective central location as the eruption progresses and the positive polarity footpoint region exhibits a more dynamic behaviour. 
The ultraviolet (UV) brightenings and extreme ultraviolet (EUV) dimmings agree well with the models in terms of their dynamics. According to our modelling results, the toroidal magnetic flux in the flux rope first rises and then decreases. In our observational analysis, we capture the descending phase of toroidal flux.
} 
{The extraction algorithm enables us to effectively study the flux rope's early dynamics and derive some of its key properties such as footpoint movement and toroidal magnetic flux. The results generally agree well with observational data.}

   \authorrunning{A.~Wagner, et al.}
   \titlerunning{The Automatic Identification and Tracking of Coronal FRs}
   \maketitle

\section{Introduction}
\label{Sect: Intro}

The largest form of solar eruptions - coronal mass ejections (CMEs) - are key drivers of space weather on Earth and other planets of our solar system. Their formation, properties and evolution are thus of major interest for the space physics community, both for accurately forecasting their space weather consequences and understanding their underlying physics.  

The common consensus is that magnetic flux ropes (FRs) are an integral part of CMEs \citep[e.g.,][]{Chen2017,Green2018}. A FR is a coherent magnetic structure where the magnetic field lines wind about a common axis. Direct observations of Earth-impacting CME FRs are usually available only at the Lagrangian point L1. Otherwise, we need to rely on occasional observational signatures at distances closer to the Sun or on remote sensing observations. The kinematic and geometric properties of CME FRs can be inferred from white-light coronagraph observations, but there are currently no suitable observations to measure the magnetic field in CME FRs remotely in a consistent manner \citep[e.g.,][]{Kilpua2019}. One can use a combination of indirect observational proxies to infer the magnetic properties of CME FRs \citep{palmerio2017}, but suitable observations or necessary signatures may not always be available. Another key issue is that the formation and early evolution of CMEs in the low corona are poorly understood. This is a severe limitation as CMEs may experience strong dynamics during this early state as they lift-off from the Sun.  

The evolution of CMEs can also be studied with numerical simulations. One of the advanced approaches is data-driven modelling, which utilizes a time series of observational data to model the time-evolution of the coronal magnetic field \citep[][]{Cheung2012, Mackay2011}. 
The modelling of the CME FR can be used to infer many of its key properties such as its magnetic twist, enclosed magnetic flux and the location of the footpoints. The analysis of footpoints does not only give information where the CME is rooted at the Sun, but their movement can reveal important aspects of a CME's early dynamics. 
The interpretation of FR observations close to the Sun is however often complicated and ambiguous and their comparison to modelling results not straightforward. Footpoints of CME FRs may manifest themselves in remote-sensing Ultraviolet (UV) and Extreme Ultraviolet (EUV) observations, both as localized dimmings and brightenings. 

The coronal dimmings refer to a localized transient decrease in the emission intensity in the vicinity of the eruption site. \citep[e.g.,][]{Thompson1998}. The CME FR’s footpoints are believed to manifest as strong (in terms of change in intensity) and localized ``core'' or ``twin dimmings'', which are found close to the eruption site  \citep[][]{Webb00, Dissauer2018}. Often also ``secondary dimmings'' are observed, which appear shallower and usually spread much further \citep[see e.g.,][]{Veronig2019, Mandrini07, Dissauer2018}. 

The dominant cause for coronal dimmings is believed to be the evacuation of plasma in the wake of the CME and/or due to its expansion, but several other sources may contribute to the dimming, such as thermal, obscuration and wave dimming \citep[e.g.,][]{Cheng2016}. Therefore, core dimmings are generally thought of as indicating the anchoring footpoints of the CME FR \citep[][]{Webb00, Dissauer2018} and the evolution of the dimming can provide unique information on the early evolution of the CME. In addition, core dimmings have also been used in the estimation of the magnetic flux enclosed in the CME \citep[][]{Mandrini2005, Webb00, Attrill2006}. 
Many effects can, however, interfere with the appearance of core dimmings that are not related to the FR footpoints, and sometimes even no dimmings are observed at all \citep[][]{Mandrini07}. 

Brightenings are intensity enhancements in UV images that typically appear as two J-shaped structures \citep[][]{Moore1995, Janvier2014}. Physically, the UV brightening signatures are interpreted as energetic particles that are accelerated at reconnection sites and subsequently interact with the chromosphere, where they cause the observed intensity enhancements. In the 3D standard model of flares as described in \cite{Janvier2014}, the hooks of these usually J-shaped areas should encircle the footpoints of the erupting structure, and thus the core dimming areas. Furthermore, according to this model, the J-ribbons should recede from each other as the eruption proceeds, together with the footpoint regions. However, reconnection in non-idealized magnetic field configurations can lead to a more complex movement, for example, instead of drifting, the footpoints can jump to a different location \citep[][]{Aulanier2019,Zemanova2019}.

In this particular work, we make use of the data-driven time-dependent magnetofrictional model (TMFM) \citep[][]{Pomoell19}, which enables us to investigate the time-evolution of FR properties during the early eruption phase. Furthermore, we perform so-called ``relaxation runs'', where we stop the driving with observational data at a given time and evolve the system with a magnetohydrodynamics (MHD) approach instead (see Sect.~\ref{Sect: MHD}).
 
In order to efficiently study the early CME FR evolution, we first develop a semi-automated tool for identifying the FR from the simulation data. One possible indicator for the identification of FRs are quasi separatrix layers (QSLs) as derived from the squashing factor $Q$ \cite{Titov2002, Demoulin1996}. QSLs have been used to aid in identifying flux ropes by some authors, such as \citet{Price19,Liu16}. Another useful FR proxy is the twist parameter $T_w$ \citep[which measures the number of turns that two infinitessimally close field lines wind about each other, see e.g.,][]{Price20,Liu16}.

An alternative twist number is also presented alongside a publicly available extraction code by \citet{Price2022}. Field line helicity is another possible proxy, for which a publicly available FR detection code exists \citep[][]{Lowder17}. For our FR extraction method, we particularly make use of $T_w$. We then apply the developed extraction procedure to National Oceanic and Atmospheric Administration (NOAA) active region AR12473, which is associated with a solar eruption on 28 Dec 2015 that produced an M1.9-class flare \citep[][]{Price20}. We derive the FR's footpoints and the toroidal magnetic flux flowing through them and analyse their temporal evolution. Furthermore, we compare the derived properties against observed core dimmings in EUV and UV brightenings. From the former, we also compute the toroidal flux evolution, which we compare against the toroidal flux evolution from the modelled FR. 

The Paper is organised as follows: In Sect.~2 we describe the used observational data as well as the used simulations. Sect.~3 presents the developed FR extraction method and derivation of FR key parameters (both from the simulation and observations). Sect.~4 shows the results derived from our analysis, which are then discussed in Sect.~5. Finally, the conclusion can be found in Sect.~6.

\section{Data and Methods}

\subsection{Observational Data}
The data-driven modeling considered in this work (detailed in Sect.~\ref{Sect: TMFM}) requires a time-sequence of photospheric vector magnetogram observations.
We use magnetograms from the Helioseismic and Magnetic Imager \citep[HMI, see][]{Couvidat16,Schou12} aboard the Solar Dynamics Observatory (SDO) spacecraft  \citep[see][]{Pesnell12}. To simulate the behaviour of active region AR12473, the simulation is run using data starting from 23 Dec 2015 23:36 UT until 2 Jan 2016 12:36 UT. The M1.9 flare occurred on 28 Dec 2015 and it peaked at 12:45 UT according to the Geostationary Operational Environmental Satellite (GOES) data.
The magnetograms are processed using the procedure presented in \cite{Price20}.

The evolution of the corona for the time and region under consideration is also investigated using observations from the Atmospheric Imaging Assembly \citep[AIA, see][]{Lemen12} instrument (also part of SDO) in the 211 \AA{} and 1600 \AA{} channels. 
The former is used for the EUV coronal dimming analysis, while the latter is used to analyse brightenings in the UV emission related to flare ribbons. 

\begin{figure}
     \centering
     \includegraphics[width=\linewidth]{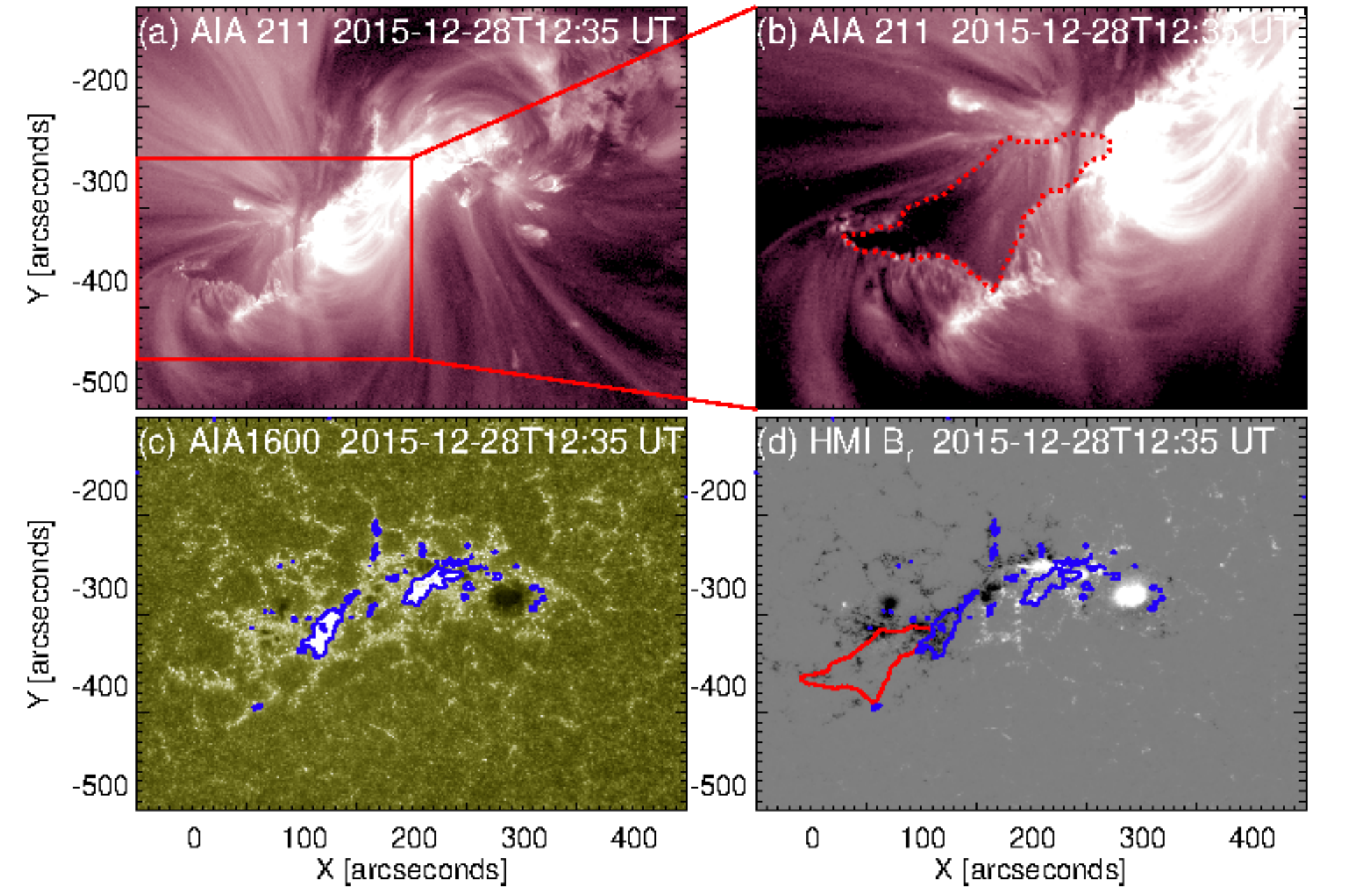}
     \caption{Core dimming and brightenings. The core dimming at the south-east part of the erupting region as observed in the AIA 211 \AA\ passband is depicted in image (a). The region bounded by the red rectangle in panel (a) is shown in panel (b). The core-dimming area is marked by the red dotted contour in panel (b). Observation of the active region in the AIA 1600 \AA\ passband is shown in image (c). Brightening associated with the flare ribbons are marked by the blue contours. The radial component of the HMI vector magnetic field is plotted in gray scale with saturation values $\pm$ 1000 G in panel (d). The blue contours in panel (d) are the same as those plotted in panel (c). The red contour in panel (d) denotes the boundary of the core dimming as marked in panel (b).}
     \label{fig: img}
\end{figure}

A showcase of the dimming region as well as the brightening regions on the 28 Dec 2015 at 12:35 UT can be seen in Fig.~\ref{fig: img}.

\subsection{Models}
\subsubsection{TMFM Model}
\label{Sect: TMFM}
To study the formation and early evolution of the FR, the data-driven and time-dependent magnetofrictional model \citep[TMFM;][]{Pomoell19} has been chosen. The simulation itself was carried out from 23 Dec 2015 at 23:36 UT to 2 Jan 2016 at 12:36 UT, with the simulation output analysed at a 6 hour cadence (starting from 24 Dec 2015 at 00:36 UT) with the simulation domain reaching a height of 200 Mm. The simulation was initialised with a potential magnetic field \citep[for details, see][]{Pomoell19} using the first magnetogram of the sequence of maps as input. For the subsequent time-steps photospheric electric field maps (electrograms) were derived from the time series of vector magnetograms \citep{Lumme17}. These electograms drive the TMFM model, which computes the evolution of the coronal magnetic field using Faraday's law, where the electric field is assumed to follow the resistive Ohm's law. In contrast to MHD approaches, the velocity in the magnetofrictional model is explicitly obtained by setting it proportional to the Lorentz force \citep{Yang1986,Pomoell19}: 

\[ \mathbf{v} = \frac{1}{\nu} \frac{\mu_0 \mathbf{J} \times \mathbf{B}}{B^{2}},\]

with $\nu$ being the magnetofrictional coefficient and $\mu_0 \mathbf{J} = \nabla \times \mathbf{B}$. With this prescription of the velocity, the system would relax towards a force-free state, if the system is no longer driven via a net Poynting flux at the boundary. However, since the system is driven in a time-dependent manner, a force-free state is not reached. We have selected a constant $\nu = 10^{-11} \frac{s}{m^2}$, except at the lower photospheric boundary, where $1/\nu$ smoothly approaches zero. The electric field at the boundary is decomposed such that it consists of an inductive and a non-inductive component. The former can be  calculated straightforwardly from Faraday's law: $\frac{\partial \mathbf{B}}{\partial t} = - \nabla \times \mathbf{E}_\mathrm{ind}$. The non-inductive component of the electric field, originating from a scalar potential and therefore not directly invertible from Faraday's law, is computed by the method described in \cite{Lumme17}, as an optimization problem. In this work, we use the optimized electrogram as determined in \cite{Price20} based on optimizing the calculated photospheric energy injection with the estimation given by the DAVE4VM method \citep[][]{Schuck08}. 

\subsubsection{Zero-Beta MHD Modelling}

\label{Sect: MHD}

The photospheric driving is a crucial element in the FR formation and destabilization. Once the FR starts rising in the corona, it is, however, likely that the lift-off continues even though the driving would cease. After this point, the photospheric boundary conditions should have only a minor effect on the modelled FR and its subsequent evolution. In this work, we switch off the driving at 29 Dec 2015 at 6:36 UT, which is approximately 18 hours after the observed M1.9 flare peak the previous day. 
The relatively long lag between the flare time and the ceasing of the driving is justified as the coronal dynamics in TMFM simulations is by construction slow, compared to typical eruptive timescales in the corona \citep[e.g.,][]{Pomoell19}.
Additionally, we evolve the resulting field configuration with a zero-beta MHD approach, as detailed in \cite{Daei23}.
On the one hand, this is done to compare with the results of the magnetofrictional model (MFM) relaxation from \cite{Price20}. They stopped the driving at 12:00 on the 29 Dec (so only few timesteps after our chosen relaxation time) and found that with MFM relaxation, the resulting FR is very similar to the fully driven TMFM FR. 
On the other hand, the zero-beta MHD relaxation may also counteract potential issues, related to the TMFM's slow evolution. 

\section{Flux Rope Extraction Algorithm and Derivation of its Key Properties}
\label{Sect: FR_algorithm}

To analyse the evolution of the FR in the simulations, it has to be first identified from a time series of data cubes containing the magnetic field information. This procedure is typically not straightforward. 
In this Section, we will first present our FR identification method and then explain how the footpoints of the FR and the enclosed magnetic flux are extracted from the simulation results. The methods, how footpoint dynamics and magnetic flux are extracted from observations are also discussed. 

\subsection{Flux Rope Extraction Method}
\label{Sect: FR}
Since FRs are commonly defined as structures, where a bundle of field lines wind about a common axis, they are expected to manifest as coherent, localised structures with high twist. Thus, our FR extraction and tracking algorithm makes use of the twist number \textit{$T_w$}, which is given by \cite{Liu16} as:

\begin{equation}
\label{equ: twist}
  T_w = \int_{L} \frac{\mu_0 J_{\parallel}}{4 \pi B} \, dl,
\end{equation}

where $J_\parallel$ is the component of the current density, parallel to the magnetic field. The integral is evaluated along a given field line $L$.
This quantity describes how many turns two infinitesimally close magnetic field lines make about each other. 

The FR extraction scheme involves the following steps: First, we identify the last time instance, where the FR is still within the boundaries of the simulation domain. The identification is done by calculating \textit{$T_w$} in a plane (cut along either x or y, cf. Fig.~\ref{fig: TMFM+MHD FR}) that the FR is expected and detected to pass through. Typically, a plane close to the polarity inversion line (PIL) of the region of interest is chosen. This procedure is done for all frames of the simulation. From these twist maps, we locate regions of high twist, which we define as $T_w > 1$ or $T_w < -1$ depending on the sign of twist of the FR. 
The extraction and tracking is then performed from the selected time instance backwards. To ascertain that the extracted high-twist region belongs to the same evolving structure, we require that two subsequent regions overlap by at least 10\%.

The tracked areas may appear quite irregular. Therefore, the shapes of the high-twist regions are reduced to circles as follows: First, the centroids of the extracted structures in the binary maps are calculated. The horizontal coordinates (i.e., either x or y - cf. Fig. \ref{fig: TMFM+MHD FR} - depending on the orientation of the twist plane) of the centroids are averaged in time, in cases where the horizontal movement does not surpass 20 Mm during the evolution (excluding outliers). In case of a significant displacement, it is fitted with a second order polynomial. In the event modelled here, the horizontal component stayed relatively stable for the TMFM case, while for the MHD simulation, it experienced significant displacement ($> 30$ Mm). The temporal evolution of the z coordinate is always fitted by a second order polynomial. This approach avoids possible abrupt jumps in the evolution of the coordinates, as well as boundary effects, which occur when the FR approaches the walls of the simulation domain (e.g., deceleration).

The determination of the circle radius is based on the area of the extracted twisted structure as follows: 
\begin{equation}
\label{equ: FR rad}
   R(t) = A(t)^{\kappa(t) - \epsilon},
\end{equation}
where $A$ is the area of the twisted region and $\epsilon$ is an empirical constant that quantifies the contribution from the highly twisted fields, which are not part of the FR itself (e.g., open field lines or field lines that connect to distinctly different regions on the photospheric boundary). The finite value of $\epsilon$ thus results in a smaller FR radius than setting $\epsilon$ to zero. $\kappa$ is constrained for each time step by finding the value that gives the largest change in the average twist, and therefore, the point where the curvature of the $\kappa$ over twist per area graph is maximal. This approach is motivated by the assumption that the FR boundaries exhibit sharp changes in the twist value. As the obtained ranges of $\kappa$ can have a large scatter, the final values are determined by a linear fit to its temporal evolution. 

As a last step, the value for $\epsilon$ is determined empirically based on the coherence and general appearance of the FR over time. To get an error estimate for the derived properties, the radius of the FR is calculated both with a non-zero, positive value as well as with setting $\epsilon$ to zero. Having the center and radius of the circle in 2D enables us to expand it to a sphere in 3D, where its surface is finally used to calculate the source points for the FR’s magnetic field lines. 
Some snapshots of the extracted FRs for AR12473 are presented in Fig.~\ref{fig: TMFM+MHD FR}.

\subsection{Footpoint Derivation from the Simulation}
\label{Sect: FP derivation}
The simulated footpoints of the FR are determined here as the intersecting points of the extracted field lines with the $z=0$ plane, the photosphere. For the evolution and morphology analysis, uniform field line sampling was used with 100 points in longitude and latitude of the spherical source for each time step, which was derived following the methodology presented in Sect.~\ref{Sect: FR}. 

To capture the dynamics of the footpoints, we calculate their respective centroid and then determine the distance of both footpoint regions from this centroid.

\subsection{Footpoint Derivation from Observations}
\label{Sect: Comp FP}
As discussed in the Introduction the hooks of the J-shaped UV brightenings (ribbons) are expected to mark the FR footpoints. For the event presented here, the brightening areas do not extend to the presumed footpoint locations, but stay rather centrally on the inner edge of the two polarity regions (cf. Fig.~\ref{fig: img}). 
To extract the brightenings from the AIA  1600 \AA{} maps, both base-difference and running-difference images have been created with a time cadence of 12 minutes. Two base images are created, as the background conditions changed significantly during the event; the average of the first two (28 Dec 2015, at 10:59 UT and 11:11 UT) and the last two (28 Dec 2015, at 13:35 UT and 13:47 UT) frames of the timeseries.

Regions that surpass a determined threshold (0.175, as found empirically to capture the main brightening area) in the difference maps are marked as brightening regions. We subsequently filter for small structures to get rid of noisy features in the resulting images. 
A binary mask outlining the extent of the ribbons over the time series is created. From this, we calculate the centroid of the ribbon structure, which is used as a reference point to estimate the movement of the brightening regions. This procedure is done for all three methods described above (base difference images using the starting frames, the final frames and the running difference images).
We then average the resulting distances from these three different methods (with respect to their respective centroid) and estimate the errors by their standard deviation. 
Finally, we want to track the movement of both brightening regions only in frames, where they appear as clear and significantly bright structures. Thus, we only consider those frames where the average intensity of the AIA 1600~\AA{} images is higher than their linear fit through the full time series. 

Another observational signature of FR footpoints are a pair of core dimming region in EUV images. For the studied event, only a single core dimming region was detected during the flare on 28 Dec 2015 between 11:49 and 14:09 UT. We use base difference images in the AIA 211~\AA\ pass band to manually extract the core dimming region as shown in panel (b) of Fig.~\ref{fig: img}, taken ten minutes before the flare peak time. The core dimming can be seen at the south-east part of the active region. However, we do not observe a co-temporal counterpart of this core dimming region in the north-west, likely because it was obscured by the overlying arcade field-lines at the north-west side of the active region. 
To capture the dimming dynamics we add a flipped version of the dimming mask to artificially create a symmetric dimming structure.
The movement of the dimming structure is then tracked (analogously to the brightenings/modelled footpoints) in reference to their respective centroid. We start our analysis with the time instance, where the dimming becomes clearly identifiable (i.e., at 12:35 UT on 28 Dec 2015).

To compare with our modelling results, the AIA images are projected to a 2D plane (assuming photospheric height levels), as our simulation domain is Cartesian. The evolution of both the EUV dimming regions as well as the UV brightenings is tracked and compared with the movement of the footpoints that are derived from the TMFM and the MHD relaxation run.

\subsection{Toroidal Flux Calculation from the Simulations}
\label{Sect: Flux}

The footpoints derived from the model serve as the basis for calculation of toroidal flux. Here, it is calculated at the base of the simulation domain, that is, at the photospheric level by summing up $\sum_{i} B_{z,i} dA$ at the footpoint locations, where $dA$ marks the area of the cells on the bottom boundary. This is done separately for both the positive and negative polarity footpoints. The resulting absolute fluxes from both regions are then averaged to give the estimation for the toroidal FR flux. Since the TMFM and MHD simulation evolve on different time scales (also with respect to the actual eruption), we use the flux evolution of the TMFM simulations as a temporal reference and define a normalised time ($\tau$) frame by setting the minimum in the TMFM toroidal flux curve to equal time 0 and the flux peak equal to time 1. The MHD normalised time is calibrated to match with the TMFM normalised time at the initiation time, while its respective flux peak is also reached at time 1. We note that while the total flux in the FR is conserved, the contributions to it from toroidal and poloidal fluxes may change along the FR. We thus average the toroidal fluxes from the positive and negative footpoints. Sampling-related uncertainties are also expected to have a small contribution.

\subsection{Toroidal Flux from Observational Data}

The toroidal flux evolution computed from the modelling are compared to imaging observations by analysing the derived core dimming region (see Sect.~\ref{Sect: Comp FP}). 

Since only one core dimming was observed (see Sect.~\ref{Sect: FP derivation})
we follow the approach by \cite{Xing20}, and compute the magnetic flux using this sole dimming region. 
Fig.~\ref{fig: img} depicts the outer boundary of the core dimming region, marked by the red dotted countour. Using the radial component ($B_r$) data from HMI vetor magnetogram, the total magnetic flux within this boundary is determined by summing up the flux ($B_rdA$; dA is the area of a pixel in the HMI data) of each enclosed pixel. To avoid the noise in magnetic field data, we include only those pixels in our analysis, where $\lvert B_r \rvert > 20$ Gauss following the threshold used by \citet{Wang_2019} and \citet{Xing20}. In order to estimate the error associated with detecting the boundary of the core dimming area, we enlarge/reduce the identified dimming boundary (while keeping the contour shapes identical) to allow a 20 \% increase/decrease in the enclosed area. This criteria of 20 \% change in boundary area is selected based on our manual inspection, so that the dimming boundary lies well within the prescribed error limit. We further estimate the associated change in underlying magnetic flux and assign this as the error in determining the toroidal flux. 

\section{Results}

\subsection{Flux rope evolution}
\label{Sect: FR Evol}

\begin{figure*}
     \centering          \includegraphics[width=\linewidth]{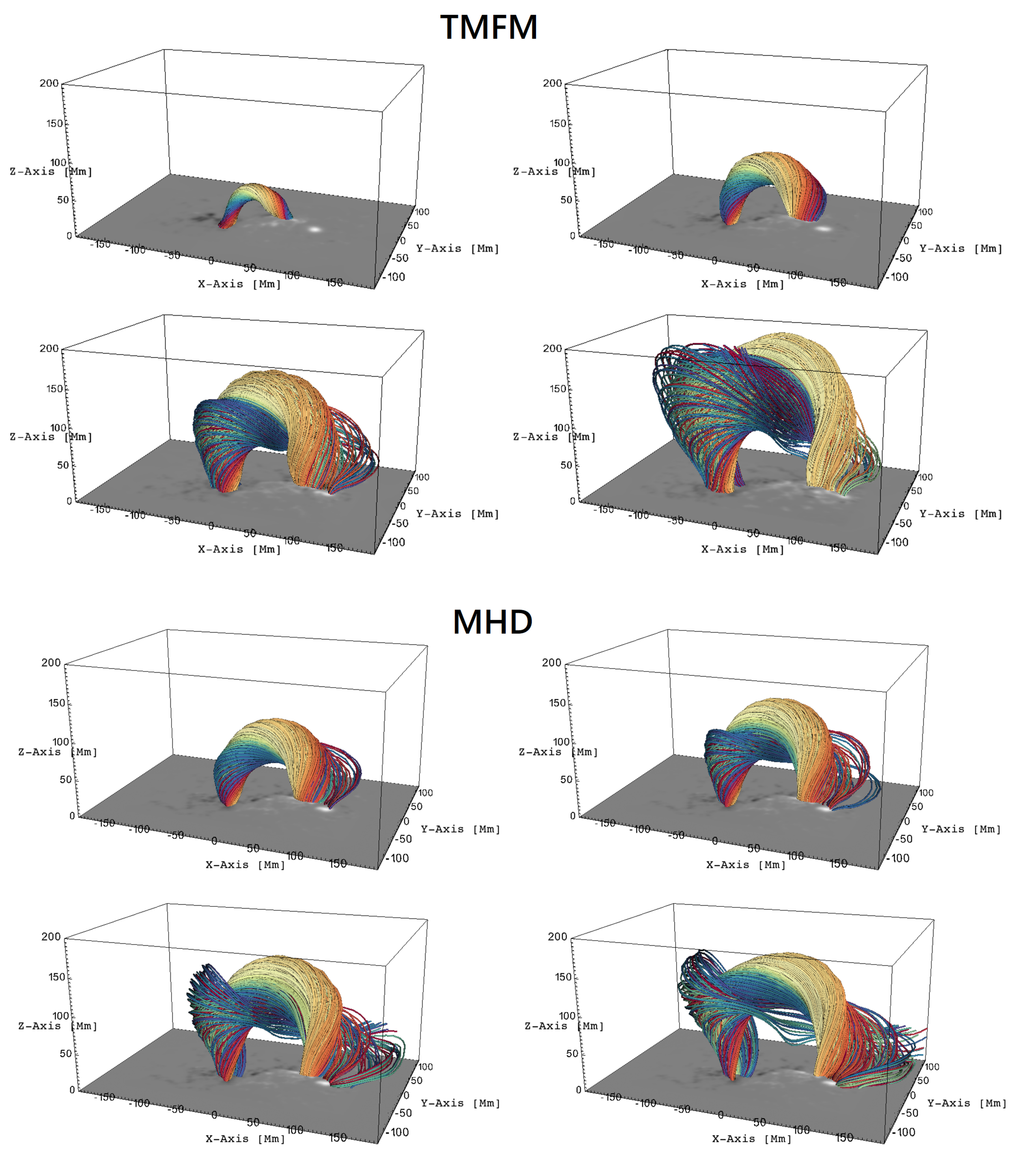}
     \caption{Snapshots showing the evolution of the FRs, extracted from the TMFM simulation (normalised simulation times w.r.t. the flux evolution, see \ref{Sect: Flux}: 0.33, 0.63, 0.92, 1.21) and the MHD relaxation (normalised times: 0.86, 1.03, 1.19, 1.36) originating from AR12473, associated with the solar eruption on 28 Dec 2015. The FR extraction method used an $\epsilon$-value of 0.03 in the TMFM case, and 0.04 for MHD (see text for details). The vertical magnetic field $B_z$ at the photospheric boundary is shown at the bottom.}
     \label{fig: TMFM+MHD FR}
\end{figure*}

\begin{figure*}
     \centering          
     \includegraphics[width=\linewidth]{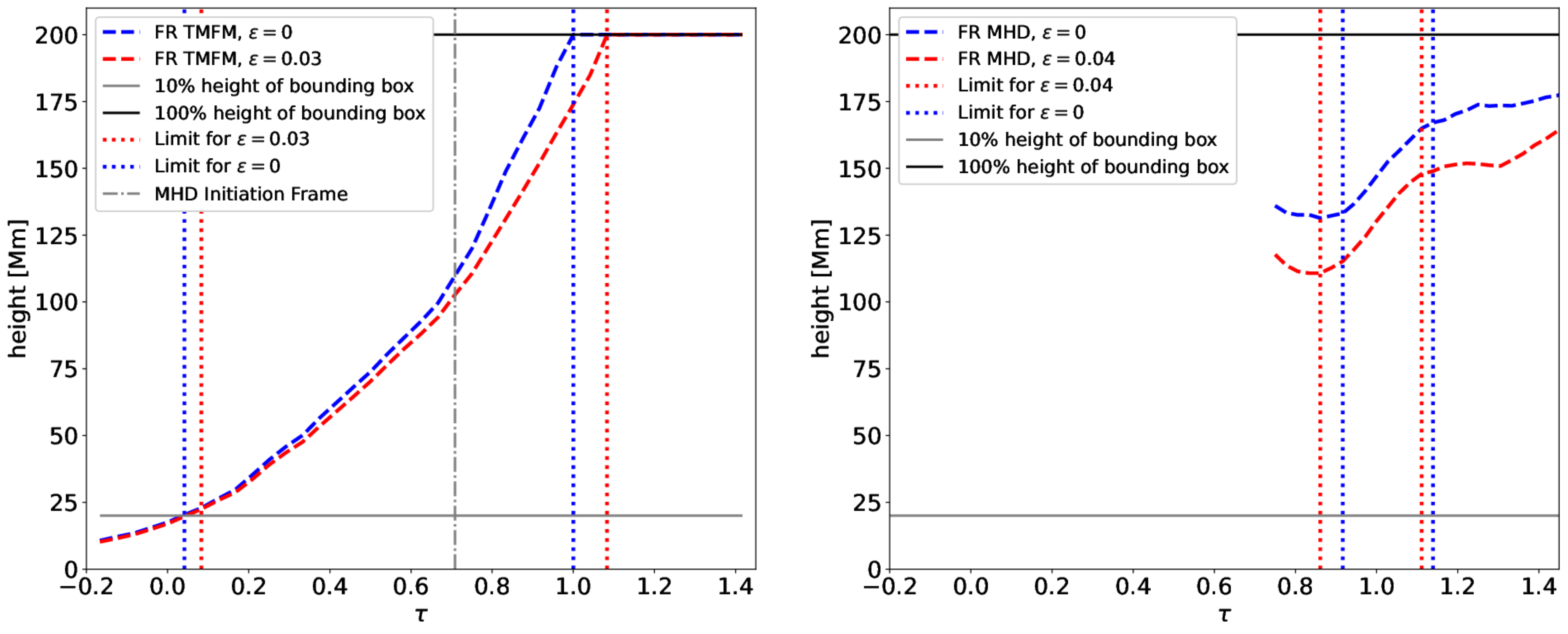}
     \caption{Evolution of the height of the apex of the FR as extracted from the TMFM simulation (left panel) and from the zero-beta MHD relaxation (right panels). The blue dashed curve shows the results when applying the extraction scheme without $\epsilon$ prescription, while the red dashed curve depicts the FR height evolution with $\epsilon = 0.03$ for the TMFM runs and $\epsilon = 0.04$ for the MHD runs (see Sect.~\ref{Sect: FR}). The x-axis depicts the time normalised to the TMFM flux evolution (\ref{Sect: Flux}). The horizontal lines show 10\% (gray) and 100\% (black) of the domain height. The dotted vertical lines indicate the start and endpoints of our analysis with (red) and without $\epsilon$-prescription (blue), as described in Sect.~\ref{Sect: FR Evol}.}
     \label{fig: Heights}
\end{figure*}

The evolution of the extracted FR in the TMFM and the zero-beta MHD simulation are visualized in Fig.~\ref{fig: TMFM+MHD FR}. The figure shows four snapshots from each model while the full evolution can be seen from the the corresponding online movies. In addition, the height of the apex of the FR as a function of normalised simulation time $\tau$ for the two simulations is shown in Fig.~\ref{fig: Heights}. We remind that the zero-beta MHD simulation is a relaxation run (see Sect.~\ref{Sect: MHD}), that was initialized with the TMFM snapshot on 29 Dec 2015 at 6:36 UT. By this time, the FR had already formed in the TMFM simulation and risen about half-way through the simulation domain. The starting height of the FR in the zero-beta MHD is $\sim 130$ Mm in the case of $\epsilon = 0$  and $\sim 110$ Mm in the case of $\epsilon = 0.04$. The differences between the TMFM initialisation height and the MHD starting height are a result of inaccuracies of the extraction algorithm for individual time steps (see a brief discussion on this in Sect.~\ref{Sect: Extraction Discussion}).

The FR computed with the TMFM (top panels of Fig.~\ref{fig: TMFM+MHD FR}) evolves very coherently and smoothly through the simulation domain until the height limit of 200 Mm. It can be visually divided into two distinct parts, illustrated with beige-reddish and blueish field lines. The structure depicted by the blueish field lines warps around the beige-reddish structure. The orientation of the field in the figure is such that the left-hand footpoint corresponds to the negative polarity one. 
The left panel of Fig.~\ref{fig: Heights} also demonstrates that while the FR evolves overall very smoothly through the low corona, the rate at which it rises increases approximately 100~Mm above the photosphere. This occurs at the simulation frame corresponding to 29 Dec 00:36 UT which is one frame (i.e., 6 hours) before the snapshot used to initiate the zero-beta MHD relaxation run.

The zero-beta MHD FR (bottom panels of Fig.~\ref{fig: TMFM+MHD FR}) has an overall quite similar appearance to the TMFM FR. It is coherent and features similar structures, with the blueish field lines, again, wrapping around the beige-red ones. The most notable differences are the tube radius, height evolution and the amount of twist, particularly of the blueish field lines. In the later stages the MHD FR appears significantly thinner, which applies to both bluish and beige-red structures. 

After initialization of the zero-beta MHD simulation, it takes some time for FR to start to rise. Then it rises at a steady rate, but slower than in the TMFM simulation, in our normalised time units. When the point is reached, where boundary effects are expected to start to affect the evolution, the rate of rise decreases (see Fig.~\ref{fig: Heights}). In contrast to the TMFM simulation, the apex of the FR does not reach the top of the simulation box but approaches it slowly. 
The slower rise and deceleration of the rise at the end of the simulation for the MHD FR may be due to the FR approaching and interacting with the upper domain boundary (for more details, see the Discussion in Sect.~\ref{Sect: Discussion}). Therefore, we exclude both the initial and end phases of the FR evolution from the subsequent analysis for the zero-beta MHD results, based on a velocity threshold (see vertical lines in Fig.~\ref{fig: Heights}). 

\subsection{Footpoint Analysis}

\begin{figure*}
     \centering
     \includegraphics[width=\linewidth]{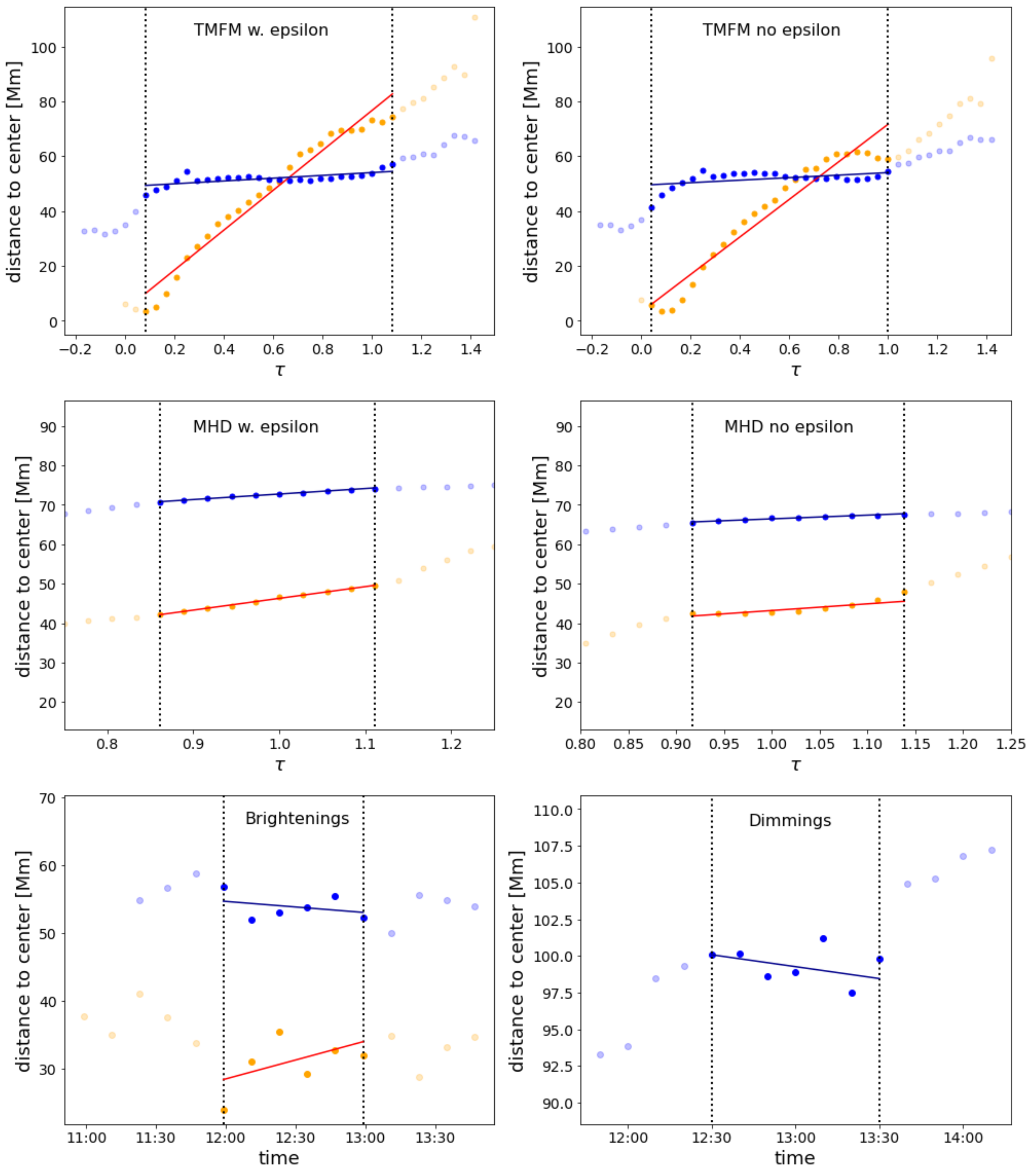}
     \caption{Distance of the footpoint/feature location from it's respective centroid for footpoints from TMFM and MHD, as well as brightenings and dimmings. The x-axis depicts the normalised time $\tau$ (see Sect.~\ref{Sect: Flux}). Orange: western footpoint (positive polarity) region, Blue: eastern footpoint (negative polarity) region. In the top two panels the pair of vertical lines shows the frames when the apex of the FR was between 10\% and 100\% of the simulation box. For the MHD results in the middle panel, we used a velocity threshold to identify relevant frames (cf. Fig.~\ref{fig: Heights}). For the brightening and dimming time window we followed the steps presented in Sect.~\ref{Sect: Comp FP}.}
     \label{fig:DistfromCenter v2}
\end{figure*}

The derived footpoint movement is detailed in Fig.~\ref{fig:DistfromCenter v2}, following the methodology introduced in Sect.~\ref{Sect: Comp FP}. The plots show the distances of both positive and negative polarity footpoints/features from their respective centroid as a function of time. 

We created linear fits through the relevant data points (as discussed in Sects~\ref{Sect: Comp FP} and \ref{Sect: FR Evol}) and calculated the average movement of the footpoints or observational feature, with respect to their centroid location. The resulting slopes of the linear fits are shown in Table \ref{table:Movement}. Note, that for the dimming, only one in the negative polarity region (eastern one) was clearly identifiable; see Sect.~\ref{Sect: Comp FP} for details. 

The faster movement of the positive-polarity footpoint regions (orange dots in Fig.~\ref{fig:DistfromCenter v2}, i.e., western footpoints) is clearly visible for both simulations and the brightenings. For both simulations the negative-polarity footpoint regions (blue dots, i.e., eastern footpoints) is relatively stable when compared to the positive-polarity footpoints. 
In both TMFM and MHD cases, the footpoints move away from their centroid - they recede from each other as the simulations progresses. Even after the FR in the MHD simulation starts to decelerate (i.e., falls below a velocity threshold), the trend for the positive-polarity footpoint continues. 

The previously described behaviour found in the simulated footpoint dynamics is less clear for the brightenings, which may be due to the limited amount of data points. The movement is generally more erratic too. The positive polarity region however exhibits both stronger dynamics as well as a positive slope, similar to simulation results. On the contrary, the negative polarity brightening shows a negative slope. The movement of the core dimming in the negative polarity region in the 211~\AA{} AIA channel agrees exceptionally well with the respective brightening results and the value of the slope is very similar, cf. Table \ref{table:Movement}. 

\begin{table*}
\begin{tabularx}{\textwidth}{|| >{\centering\arraybackslash}X 
| >{\centering\arraybackslash}X 
| >{\centering\arraybackslash}X ||} \hline \label{table:Movement}
Data source & Slope (neg. pol. footpoint) & 
Slope (pos. pol. footpoint)\\
\hline\hline
TMFM FR & $0.198 \pm 0.006 $  & $2.878 \pm 0.025 $ \\
\hline
MHD FR & $0.363 \pm 0.069 $ & $0.722 \pm 0.196 $ \\
\hline
Brightenings & $-0.274 \pm 0.133 $ & $0.929 \pm 0.235 $  \\
\hline
Dimmings & $-0.270$ & - \\
\hline
\end{tabularx}

\caption{Slopes derived from linear fits from Fig.~\ref{fig:DistfromCenter v2}. The slopes quantify the strength of movement with respect to the centroid location of the respective feature. Positive slopes correspond to movement away from the centroid, while the feature moves towards its centroid if the slope is negative. The slopes for the modelling data are in units of megametres per timestep, while the slopes for the observational data are in units of megametres per 10 minutes.}
\end{table*}

\subsection{Toroidal Flux Analysis}

\begin{figure*}
     \centering
     \includegraphics[width=\linewidth]{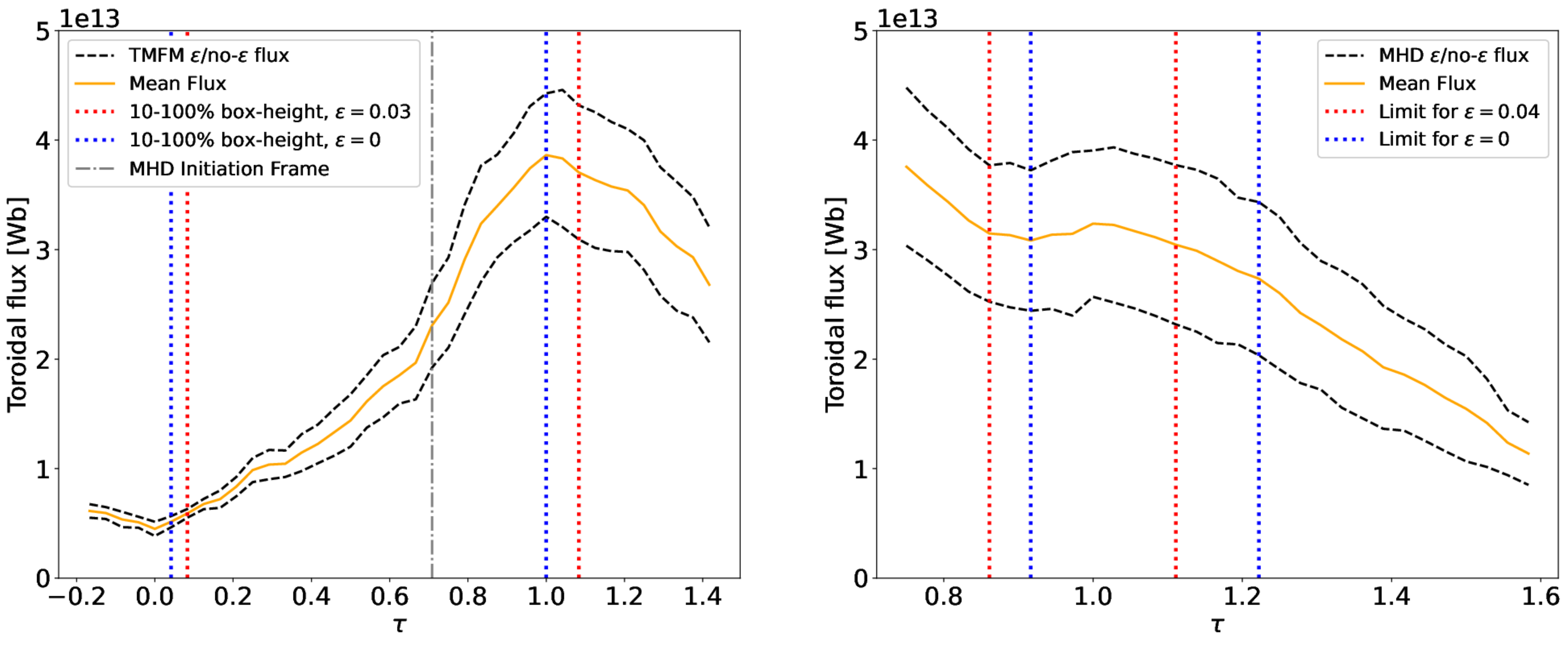}
     \caption{FR flux in the TMFM simulation as well as the zero-beta MHD simulation for the same event, with the normalised time $\tau$ on the x-axis (see Sect.~\ref{Sect: Flux}). The red and blue pairs of dashed lines indicate the start and endpoints of the most relevant frames for the simulated FRs with and without taking $\epsilon$ into account, respectively. The dashed gray line in the TMFM plot indicates the initiation time for the zero-beta MHD simulation.}
    \label{fig:FR flux}
\end{figure*}

The results of the computation of the toroidal magnetic flux enclosed in the FR (for details, see Sect.~\ref{Sect: Flux}) is presented in Fig.~\ref{fig:FR flux}. Similarly to the footpoint analysis, for the TMFM FR the time interval is shown when the FR height is between 10 and 100 \% of the domain height, while for the MHD FR, a velocity threshold of 1 Mm per time step has to be reached. 

The TMFM FR shows a nearly monotonous increase in toroidal flux until the FR approaches the top boundary of the simulation domain. After reaching the top (or shortly before in the case of the $\epsilon = 0.03$ FR), the flux starts declining. 
For the MHD FR the flux first decreases before it increases towards the peak value (note, that we consider here only times where the velocity limit was satisfied). After reaching this maximum, the flux declines steadily. Contrary to the TMFM FR, this decline already sets in, when boundary conditions should not yet play a role in the FR's evolution. 

\begin{figure*}
     \centering     \includegraphics[width=\linewidth]{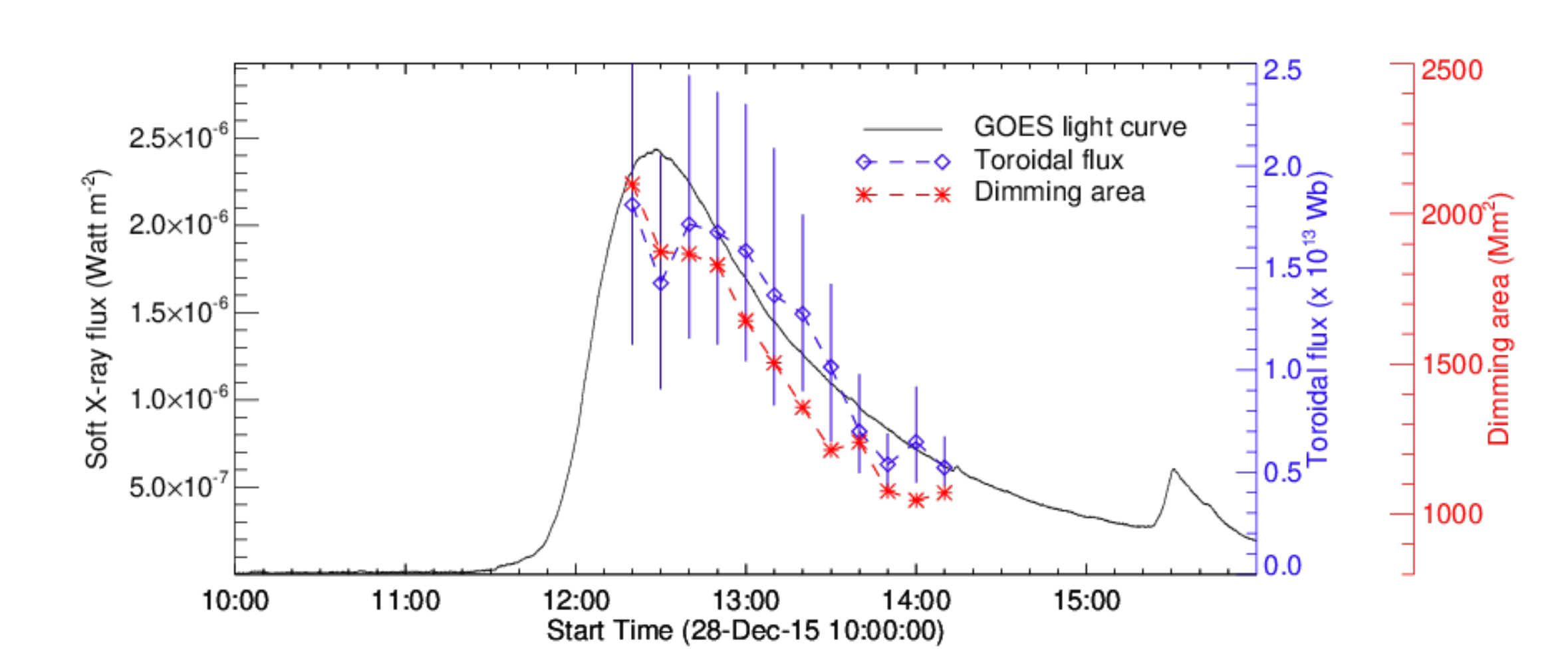}
     \caption{Temporal evolution of the toroidal flux and the core-dimming area. The black curve denotes the co-temporal evolution of the GOES soft X-ray flux. }
     \label{fig:FR dimm flux}
\end{figure*}

The flux derived from the core dimming observation in 211~\AA{} is shown in Fig.~\ref{fig:FR dimm flux} together with the dimming area and the GOES light curve. The rising phase observed in the model results could not be captured, but a clear declining phase matches especially well with the MHD results. \\

\section{Discussion}
\label{Sect: Discussion}
\subsection{FR Extraction Discussion}
\label{Sect: Extraction Discussion}
The developed semi-automated extraction of the FR from the simulation data is highly beneficial for studies of a FR's early dynamics and evolution. The method uses a twist threshold (here twist number $|T_w| > 1$), a requirement of coherency of the rising structure and constraints of a circular cross-section in a plane close to the polarity inversion line. 
The radius of the circle was determined by finding the value that gives the largest change in the average twist.

There are, however, some steps in the extraction procedure that may affect the outcome of the analysis. Firstly, the twist threshold may influence the extent of the extracted FR and therefore the centroid of the FR structure. The radius of the source sphere itself is, however, only weakly dependent on the twist threshold, as the $\kappa$ (Eq.~\ref{equ: FR rad}) is ultimately responsible for setting the source radius. $\kappa$, thus counteracts possible non-optimal thresholds for the twist map, as long as they are in a reasonable range. Secondly, FR cross-sections may also deviate significantly from circular shapes \citep[][]{Price20,Kilpua2021}. Furthermore, since we use fitting procedures over the whole simulation time to smooth the FR's centroid and radius evolution, individual frames may not completely capture the FR at a given time instance.
This explains why in our simulation the MHD FR at the initiation time had quite different values of toroidal flux (cf. Fig.~\ref{fig:FR flux}) as well as height values (cf. Fig.~\ref{fig: Heights}), compared to the TMFM FR.

Furthermore, we use here a version of the twist number ($T_w$) that describes the number of turns a field line takes around an infinitesimally close field line  \citep{Liu16}. This is therefore different from the general definition of the FR as a structure, where field lines wind about a common axis. In this regard a more accurate approach would have been to use the twist number $T_g$, which is defined as the number of turns a field line takes about a defined (FR) axis \citep[e.g.,][]{Liu16,Price2022,Berger2006,Guo17}.
The computation of the FR axis is however non-trivial and the procedure requires the knowledge of $T_w$ \citep[][]{Price2022}. Therefore, we chose that the most feasible approach for automated flux rope identification is to use $T_w$ instead of $T_g$.

\begin{figure*}
    \centering    \includegraphics[width=\linewidth]{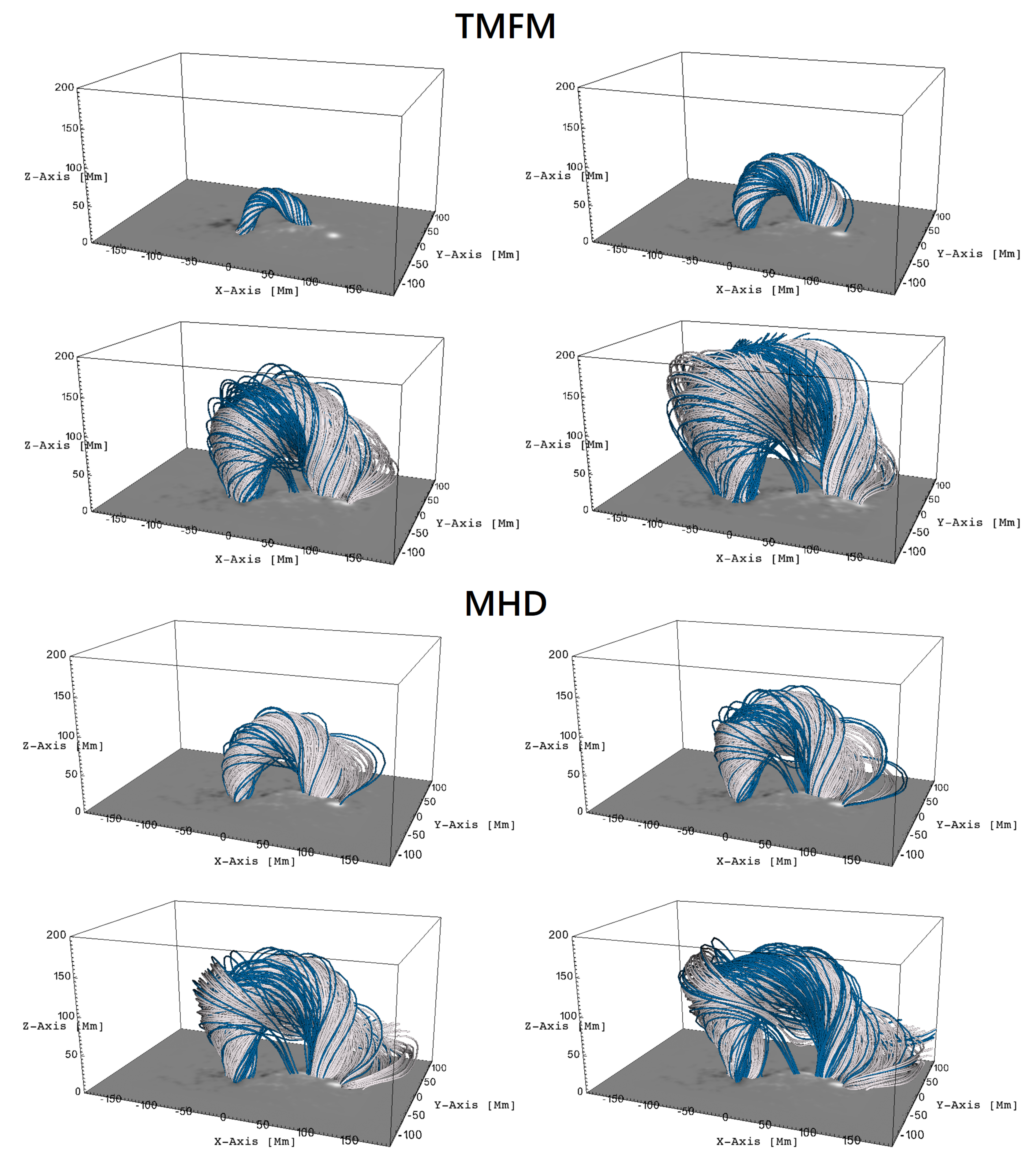}
    \caption{Showcase of FR field lines from the extraction method with $\epsilon$ prescription (white) and without (blue) for the TMFM and the MHD relaxation simulation. The frames are the same as in Fig.~\ref{fig: TMFM+MHD FR}}
    \label{fig: Epsilon}
\end{figure*}

We incorporated a correction parameter ($\epsilon$) to reduce the amount of highly twisted surrounding field that has distinctly different topology and/or connectivity at the photospheric level. While the $\epsilon$ parameter may be very effective in removing unwanted field lines, it also comes with trade-offs: The surrounding, non-FR field lines are most of the time not equally distributed around the FR boundaries, and thus reducing the circular cross section of the twisted structure cuts out actual FR field lines too. This is the reason why we computed the height, footpoints and flux evolutions both with and without the $\epsilon$ prescription. This also aids to obtain an error estimation. We used a fixed value for this correction parameter, but it could be modified to incorporate a time-dependence in the future.
A showcase of the effect of applying $\epsilon$ to the FR can be seen in Fig.~\ref{fig: Epsilon}. The removal of field lines that connect to different regions in the photosphere than those in the main FR, can be seen clearest in the later stages of the simulations (bottom panels for both TMFM and MHD). We note that a significant amount of excluded blue field lines blend in well with the main FR. 

The results of the footpoint derivation and especially the flux calculation may also be affected by the inclusion of magnetic field lines that exit the side walls of the simulation domain. For these simulations, their contribution was, however, minimal (some can be seen in the bottom panels for the MHD run in Fig.~\ref{fig: TMFM+MHD FR}). 

\subsection{Modelling Discussion}
In this work, we applied the extraction method for two simulations, 1) a TMFM simulation and 2) a relaxation run, using a zero-beta MHD model \citep{Daei23} that was initiated using a TMFM snapshot approximately halfway through the TMFM simulation without any additional driving after this point. The exact relaxation time of 29 Dec 2015 at 6:36 UT yielded an erupting FR, consistent with the magnetofrictional relaxation carried out by \cite{Price20}. 

The extracted FRs generally agree well with the common perception of a magnetic FR: They feature a coherent and gradually rising structure, containing twisted field lines winding about a shared axis. We found that the MHD FR deviates a bit more from its original coherence in the later stages than the TMFM FR (cf. Fig.~\ref{fig: TMFM+MHD FR}). Furthermore, the rise of the MHD FR decelerates and it remains within the domain for the whole simulation duration. It has to be noted, that this is most likely a boundary effect and not the FR finding an actual new equilibrium.
We therefore restricted our analysis to those frames, where the MHD FR was rising sufficiently fast (at a velocity of at least 1 Mm per time step). This requirement also excluded the very early phase, where the FR's instability was slowly building up.
The TMFM FR evolution was limited by reaching 10\% (lower limit) and 100\% (upper limit) height of the simulation domain. 
Within their respective boundaries, the appearance and dynamics of the TMFM and MHD FRs seem to agree well with each other, both in terms of general morphology and temporal evolution. 

\subsection{Footpoints and Toroidal Flux}
The footpoint movement, as deduced from the modelling and the observational analysis (i.e., from the dimming and brightenings) generally agree well with each other. There are, however, some differences in the dynamics, for example, how fast footpoints recede from the common centroid. This is presumably due to the fact that the observational proxies cover only some particular phase of the FR evolution, while in the simulation, the FR can be tracked until it reaches the top of the domain or boundary effects start to interfere. Furthermore, the movement of the negative polarity footpoint region in observations is directed towards their respective centroid, contrary to the modelling results. We, however, remind that the movement of the negative polarity footpoint was quite minor when compared to the evolution of the positive polarity region and thus no strong conclusions can be drawn. 
The observational features may also partly capture other parts of the FR system than the footpoints. As discussed in the Introduction (\ref{Sect: Intro}), the UV brightenings can be complex and reflect interplay between the FR fields and surrounding arcade fields. Although the brightenings' evolution should correlate with the movement of the footpoints during the eruption, they give information on the current sheets related to the erupting structure, rather than on the footpoints directly. The generally different underlying effects can also be understood, when observing that the relevant time windows for the brightenings and dimming are shifted by 30 minutes between each other. We note, that the slopes for the negative polarity regions derived from dimming and brightenings match exceptionally well, suggesting that they both have captured the same kind of moving structure. The trend of the footpoints receding from the centroid is generally evident both from the simulated and the observed data in the very dynamic, positive polarity footpoint region. The dynamics of the positive-polarity footpoint region are present both in the simulation and brightenings. Its behaviour could not be reliably captured by the dimming analysis, however, as we do not observe any dimmings in this particular region. The derived slopes (see \ref{table:Movement}) summarise these trends, though their magnitude differs notably between the simulation and observational features due to different time scales as discussed above. 

We find that for the simulated FRs within their respective restricted time windows, the toroidal flux first increases steadily. After reaching its peak, we observe a declining trend, which is especially pronounced in the MHD case. The decline in toroidal flux sets in rather late for the TMFM FR, when it approaches the domain boundary. We cannot, however, draw strong conclusions after the time when the FR starts exiting the simulation domain as this directly reduces the flux.

The two-phase evolution of toroidal flux that we see in simulations is consistent with the observational investigation by \cite{Xing20}. In our case, we could only capture the declining phase from observations, that is, from the core-dimming analysis (see Fig.~\ref{fig:FR dimm flux}). This could also be due to the fact that the dimming evolution was not that clearly trackable in the beginning phase of the flare. The two-phase evolution of the toroidal flux is speculated by \cite{Xing20} to be caused by the flux first increasing as the FR quickly forms by reconnection with the overlying field. Subsequently, the decrease sets in due to reconnection between the twisted fields that are within the FR, converting toroidal to poloidal fields. 

\section{Conclusion}
We developed a semi-automatic extraction algorithm to track eruptive coronal FRs in data-driven simulations.
The method was applied to the active region AR12473, using both the time-dependent magnetofrictional method (TMFM) and a zero-beta MHD relaxation approach. Both simulations resulted in an erupting FR.  
The evolution of toroidal magnetic flux and the footpoint movement was determined from the simulations and compared with the coronal brightenings and dimmings. We find that during the eruption process the modelled FRs' footpoints as well as the brightenings and dimmings showcase a movement away from their respective central location, which is stronger for the positive-polarity region. The resulting trends are summarised in Table \ref{table:Movement}. 
The derived toroidal magnetic FR flux through the modelled footpoints shows a two-phase evolution: a rising and a declining phase, which is most clear in the MHD case. Our observational investigation shows mostly a decline in toroidal flux, though we may not be capturing the build up of flux, as it may have happened before the dimmings were well-observable. 

Our study shows that the developed algorithm can successfully capture the early dynamics of the FR and derive the evolution of some of its key properties. 
This tracking method also facilitates statistical analysis of simulated FRs and tracking the evolution of their parameters. 
Refining the FR extraction method, and applying it to investigate other aspects of FR eruptions could be an avenue for future research as well. 

\begin{acknowledgements}
SDO data are courtesy of NASA/SDO and the AIA and HMI science teams. This research has made use of NASA’s Astrophysics Data System. Furthermore, to compute and visualize the flux ropes, the open source python visualization tool VisIt \cite{HPV:VisIt} was used, as well as the PyVista package for multiple applications within this paper \cite{Sullivan19}. 
This project has received funding from the European Union’s Horizon 2020 research and innovation programme under the Marie Skłodowska-Curie grant agreement No 955620. SP acknowledges support from the projects
C14/19/089  (C1 project Internal Funds KU Leuven), G.0B58.23N (FWO-Vlaanderen), SIDC Data Exploitation (ESA Prodex-12), and Belspo project B2/191/P1/SWiM.
 E.K.J.K., A.K and F.D acknowledge the ERC under the European Union's Horizon 2020 Research and Innovation Programme Project SolMAG 724391, and Academy of Finland Project 310445. R.S acknowledges support from the project EFESIS (Exploring the Formation, Evolution and Space-weather Impact of Sheath-regions) under the Academy of Finland Grant 350015. J.P and F.D acknowledge Academy of Finland project 343581. All UH authors acknowledge the Finnish Centre of Excellence in Research of Sustainable Space (Academy of Finland grant numbers 312390, 312357, 312351 and 336809).
\end{acknowledgements}
\bibliographystyle{aa}
\bibliography{output}
\end{document}